\documentclass[11pt]{article}
\usepackage{graphicx}
\newcommand{\bda}{\begin{\displaymath}\begin{array}{rl}}
\newcommand{\eda}{\end{array}\end{displaymath}}
\newcommand{\be}{\begin{equation}}
\newcommand{\ee}{\end{equation}}
\newcommand{\bdm}{\begin{displaymath}}
\newcommand{\edm}{\end{displaymath}}
\newcommand{\bea}{\begin{eqnarray}}
\newcommand{\eea}{\end{eqnarray}}

\begin{document}

~\vspace{1cm}
\begin{flushright}{\it Dedicated to Professor Mihai Gavril\u a's $80^{th}$ Anniversary} \end{flushright}

~\vspace{0.4cm}

\begin{center} 
{\bf\Large NEW PERTURBATION EXPANSIONS \\\vspace{1mm}IN QUANTUM CHROMODYNAMICS \\\vspace{2mm} AND THE DETERMINATION OF $\alpha_s$}

~\vspace{1cm}

Irinel Caprini$^a$,  Jan Fischer$^b$

~\vspace{0.4cm}

$^a$National Institute of Physics and Nuclear Engineering, \\POB MG 6,
Bucharest, R-077125 Romania \vspace{1mm}\\
$^b$Institute of Physics, Academy of Sciences of the Czech Republic, \\
CZ-182 21  Prague 8, Czech Republic

\end{center}

~\vspace{1cm}

\begin{abstract}  We consider  a new class of perturbation expansions, which incorporate in a systematic way the available information about the  divergent character of the perturbation series in QCD. The new expansion  functions, which replace the powers of the coupling,  are defined by the analytic continuation in the Borel  plane, achieved through an optimal conformal mapping.  We consider several possibilities of implementing the known  behaviour of the Borel transform near the leading singularities in the complex  plane and using the corresponding expansions. These expansions have the same asymptotic convergence rate, but  differ at low orders. We show that the  new expansions allow a precise determination of  the strong coupling  $\alpha_s$ from the hadronic decays of the $\tau$ lepton.\end{abstract}

\section{INTRODUCTION \label{intro}}

As it is known, in 1952   Dyson obtained the famous result \cite{Dyson} that the renormalized perturbation series in Quantum Electrodynamics is divergent.  During the subsequent decades, similar  results have been obtained \cite{Lautrup}-\cite{Broad} for most of the physically interesting  field theories, including Quantum Chromodynamcs, the modern theory of strong interactions. Dyson's proposal  to regard a divergent series as asymptotic to the function sought for is nowadays generally adopted.  This set a challenge for a radical reformulation of perturbation  theory.  The Feynman diagrams  and renormalization theory yield, at least in principle,  the values of all the expansion coefficients. They  can tell us whether the series is convergent or not, but what we want to 
know is under what conditions the physical function can be determined from them. If the series  were  convergent,  the knowledge of all the expansion coefficients would uniquely determine the expanded function.   
On the other hand, there are infinitely many functions having the same 
asymptotic expansion.  A crucial task is to
find effective additional inputs that would be able to reduce or, if possible,
remove the ambiguity. For instance, general properties such as causality and unitarity  may be useful.  For a recent discussion of the the ambiguity of field correlators represented  by asymptotic perturbation expansions  see \cite{CaFiVr}.

Along these lines, a possibility of exploiting better the low order  expansion coefficients known at present from Feynman diagram is to combine them  with  the available information on the large-order behaviour of the expansion.  Such an attempt was made in  \cite{CaFi}, where a  new expansion was defined by  the  analytic continuation in the Borel complex  plane achieved by a  conformal mapping\footnote{For QCD, the idea of a conformal mapping in the Borel plane  was suggested in \cite{Muel1} and was applied in several works (for references see \cite{CaFi}).}. Specifically, we use an optimal conformal mapping, as defined in \cite{CiFi}.    As shown in \cite{CaFi, CaFi1, CaFi2}, the  new perturbation expansion   separates the divergent character of the series from a part on the intrinsic ambiguity  of the  perturbation theory, which is  generated by the infrared regions of the Feynman diagrams. This ambiguity is  solved by choosing a prescription included in the definition of the expansion functions.

In the present paper we discuss these ideas using for illustration the so-called Adler function \cite{Adler}. This function is of interest in QCD for the determination of the strong coupling $\alpha_s$ at a relatively low scale, the mass of the $\tau$ lepton \cite{BNP92}-\cite{Bethke}. 
In section \ref{sec:stand} we briefly review the standard perturbation expansion of the Adler function in massles QCD. In  section \ref{sec:new} we present  the new perturbation expansion proposed in \cite{CaFi} and in section \ref{sec:gen} we investigate a more general class of expansions functions, which have the same asymptotic convergence rate but differ at low orders.  The convergence of the new expansions is checked using  a realistic model proposed recently in the literature \cite{BeJa}. In section \ref{sec:alphas} we  briefly discuss the  determination of $\alpha_s(m_\tau^2)$, confirming the precise value reported recently in \cite{CaFi3}.

\vskip0.5cm

\section{STANDARD PERTURBATION EXPANSION  \label{sec:stand}}

The  Adler function  \cite{Adler} in massless QCD is the derivative 
\begin{equation}
D(s) =- s\frac{{\rm d}\Pi(s)}{{\rm d}s},
\label{D}
\end{equation}
where $\Pi(s)$ is the invariant amplitude of the correlator
\begin{equation} 
i \int{\rm d}^{4}x e^{iq.x}<0|T\{V_{\mu}(x)V_{\nu}(0)^{\dag}\}|0>
= (q_{\mu}q_{\nu}-g_{\mu\nu}q^{2})\Pi(s).
\label{correl}
\end{equation}
Here $s=q^2$ is the energy squared  and $V_\mu$ is the vector current for light quarks. 
 
 In perturbative QCD,  the reduced Adler function $\hat D(s)=D(s)-1$  is written  formally as  the renormalization-group improved series  
  \begin{equation}\label{DCI} 
\hat D(s) =   \sum_{n\ge 1}  K_{n}\,  (a_s(s))^n\,, \end{equation} 
where $a_s(s) \equiv \alpha_s(s)/\pi$ is the running coupling.
  Explicit calculations of  Feynman diagrams in  $\overline{\rm MS}$ renormalization scheme with three flavours ($n_f=3$), predict the first four coefficients \cite{KaSt}-\cite{BCK08}
\be\label{Ki}
K_1=1,~~ K_2= 1.639,~~  K_3=6.371,~~ K_4=49.076,
\ee
while an estimate of the next coefficient 
\be\label{K5}
K_5 \approx 283
\ee
is available  \cite{BeJa}.

As discussed above, the series  (\ref{DCI}) is divergent, the coefficients displaying  a factorial asymptotic increase, $K_n\sim n!$ \cite{Beneke1993, Broad}.  In writing  (\ref{DCI})  we follow the convention often adopted in physical papers, writing the sign of equality even if the series on the right hand side is divergent and the equality is impossible.  Analogous series in the present paper  are understood in the same sense. 

The running coupling  $a_s(s)$  is obtained by setting the scale $\mu^2=s$ in the renormalized coupling  $a_s(\mu^2)=\alpha_s(\mu^2)/\pi$, which  satisfies the renormalization group equation
\be\label{reneq}
\mu^2 \frac{{\rm d} a_s(\mu^2)}{{\rm d}\mu^2}= \beta(a_s(\mu^2))\equiv -\sum_j \beta_j (a_s(\mu^2))^{j+2}
\ee
 The first two coefficients  of the $\beta$ function are universal (scheme-independent), and the next two were calculated in the  $\overline{\rm MS}$ scheme with $n_f=3$ to four loops  \cite {LaRi, Czakon}. Therefore, the known coefficients are:
\be\label{betai}
\beta_0=9/4,~~\beta_1=4,~~\beta_2=10.06,~~\beta_3=47.23.
\ee 

\vskip0.5cm

\section{NEW PERTURBATION EXPANSIONS  \label{sec:new}}
To define a new series, we start from the Borel transform $B(u)$  of the Adler function, defined by the power series
\begin{equation}\label{B}  B(u )=\sum\limits_{n=0}^\infty b_n u^n\,, 
\end{equation} 
 with  $b_n$ defined in terms of the perturbative coefficients $K_n$ by: 
\begin{equation}\label{bn}
b_n=\frac{K_{n+1}}{\beta_0^n \,n!}\,,\quad n\ge 0. \end{equation}  
 According to present knowledge, the function $B(u)$ has   branch
point  singularities  in the $u$-plane, along the negative
axis - the ultraviolet (UV) renormalons - and along the positive axis - the  infrared
 (IR) renormalons \cite{Beneke1993, Broad}. Specifically, the branch cuts are situated  along the rays $ u \leq -1$ and $u \geq 2$.  The nature of the first
branch points  was established in \cite{Muel} and in \cite{BBK} (see also \cite{BeJa}). Thus, near the first branch points, {\em i.e.} for $u\sim -1$ and $u\sim 2$, respectively,  $B(u)$  behaves as
\be\label{branch}
B(u)\sim \frac{r_1}{(1+u)^{\gamma_1}},\quad B(u)\sim \frac{r_2}{(1-u/2)^{\gamma_2}},
\ee
where the residues $r_1$ and $r_2$ are not known, but the exponents $\gamma_1$ and $\gamma_2$ can be calculated using renormalization group invariance \cite{Muel, BBK, BeJa}, and have the values
\be\label{gammai}
\gamma_1= 1.21,\quad \quad \gamma_2=2.58.
\ee 

The series (\ref{DCI}) can be formally written as  the  Borel-Laplace transform
\begin{equation}\label{Laplace}
  \hat D(s)=\frac{1}{\beta_0}\,\int\limits_0^\infty\!{\rm e}^{-u/(\beta_0 a_s(s))} \, B(u)\,{\rm d}u. \end{equation}
Actually, due to the singularities of $B(u)$  along the positive axis, the integral (\ref{Laplace}) does not exist. The ambiguity in the choice of prescription is often used as a measure of the uncertainty of the calculations in perturbative QCD.
 It is convenient to define the integral by the Principal Value (PV) prescription:
\be\label{Dpv}
\hat D(s)\equiv\frac{1}{\beta_0}\,{\rm PV}\int\limits_0^\infty\!{\rm e}^{-u/(\beta_0 a_s(s))} \, B(u)\,{\rm d}u\,,
\ee
where
\bea\label{Pv}
{\rm PV} \int\limits_0^\infty f(u)\,{\rm d}u  \equiv  \lim_{ \epsilon \to 0}\, \frac{1}{2}\int\limits_0^\infty \left[ f(u+i \epsilon)\,{\rm d}u + f(u-i \epsilon)\,{\rm d}u\right]. \nonumber
\eea
As discussed in \cite{CaNe} the PV  prescription  is the best choice  if one wants to preserve as much as possible  the analyticity properties of the correlators in the  $s$-plane, which are connected with causality and unitarity.

In order to define a new perturbative expansion of the Adler function we shall apply the method of conformal mappings \cite{CiFi}.
This method  is not applicable to the series (\ref{DCI}), because 
${\hat D}(s)$ (regarded as a function of $a_{s}(s)$) is singular at the point of 
expansion $a_{s}(s)=0$. The method can, on the other hand, be applied to 
(\ref{B}), because $B(u)$ is holomorphic at $u=0$. 

We note that  the expansion (\ref{B})  converges only in the disk $|u|<1$. A series  with a larger domain of convergence can be obtained by expanding $B(u)$ in powers of a new variable.   As demonstrated in \cite{CiFi}, the optimal variable, which leads to the best asymptotic convergence rate,  coincides with the function that performs the conformal mapping of the whole analyticity domain of the expanded function onto a disk in the new complex plane.

It is generally assumed that $B(u)$ is analytic in the $u$-plane cut along the real axis  for $u\ge 2$ and  $u\le -1$ (for more remarks see \cite{CaFiVr, CaFi3}). Then the optimal variable  defined in \cite{CiFi}  reads \cite{CaFi}
\be\label{w}
w(u)=\frac{\sqrt{1+u}-\sqrt{1-u/2}}{\sqrt{1+u}+\sqrt{1-u/2}}.
\ee
This function maps the $u$-plane cut for $u\ge 2$ and $u\le -1$ onto the unit disk $|w|<1$  in the complex plane $w=w(u)$, such that $w(0)=0$,  $w(2)=1$ and  $w(-1)=-1$. Then,   the expansion
 \be\label{Bw0}
B(u)=\sum_{n\ge 0} d_n \,w^n
\ee
converges in the whole disk $|w|<1$ \cite{CiFi, CaFi}.  Moreover, as shown in \cite{CiFi}, the expansion (\ref{Bw0}) has the best asymptotic rate of convergence compared to all the expansions of the function $B(u)$ in powers of other variables. 

The series (\ref{Bw0}) can be used to define an alternative  expansion of $\hat D(s)$. This is obtained formally by inserting (\ref{Bw0}) into  (\ref{Dpv}) and interchanging the order of summation and integration. Thus, we adopt the modified expansion,  defined as \cite{CaFi}-\cite{CaFi2}
\be\label{DnewCI0}
\hat D(s)=\sum\limits_{n\ge 0} d_n W_n(s),
\ee
in terms of the expansion functions
\be\label{Wn0}
W_n(s)=\frac{1}{\beta_0}{\rm PV} \int\limits_0^\infty\!{\rm e}^{-u/(\beta_0 a_s(s))} \, w^n\,{\rm d}u\,.
\ee

We emphasize that the expansion (\ref{Bw0}) exploits only  the location of 
 the leading  singularities in the Borel plane. The series is expected to describe also  the nature of the singularities  if a large number of terms is used.   However, since the behaviour near the first singularities  is known, cf. eq. (\ref{branch}), it is convenient to  incorporate it  explicitly.  This is achieved, for instance, by expanding the product $(1+w)^{2\gamma_1}(1-w)^{2\gamma_2}\, B(u)$ in powers of the variable $w$:
 \be\label{Bwgamma}
(1+w)^{2\gamma_1}(1-w)^{2\gamma_2} B(u)= \sum\limits_{n\ge 0} c_n \,w^n.
\ee
Here we took into account the fact that from the expression  (\ref{w}) of $w=w(u)$ it follows that  $(1-u/2) \sim (1+w)^2$ and $(1+u) \sim (1-w)^2$  near the points $w=-1$ and  $w=1$, respectively.

The expansion (\ref{Bwgamma})  converges in the whole disk  $|w|<1$, {\em i.e.} in the whole cut complex $u$-plane. Moreover, since the singular behaviour of $B(u)$ at the first branch points is compensated by the first factors in  (\ref{Bwgamma}), the series is expected to converge faster than (\ref{Bw0}). Also, the behaviour near the first singularities holds even for truncated expansions, which are used in practice. This suggests  the definition of the new expansion
\be\label{DnewCI}
\hat D(s)=\sum\limits_{n\ge 0} c_n {\cal W}_n(s), 
\ee
where the expansion functions are 
\be\label{Wn}
{\cal W}_n(s)=\frac{1}{\beta_0}{\rm PV} \int\limits_0^\infty\!{\rm e}^{-u/(\beta_0 a_s(s))} \,  \frac{w^n}{(1+w)^{2\gamma_1}(1-w)^{2\gamma_2}}\,{\rm d}u\,,
\ee
with $w=w(u)$ defined in (\ref{w}). Actually, as discussed in \cite{CaFi, CaFi3},  the inclusion of the explicit behaviour near the leading singularities is not unique. The problem will be discussed briefly in the next section.

The expansions (\ref{DnewCI0}) and  (\ref{DnewCI}) reproduce the coefficients $K_n$ of the usual expansion (\ref{DCI}), when the  functions (\ref{Wn0}) and (\ref{Wn}) are expanded  in powers of the coupling. As shown in \cite{CaFi2}, the new expansion functions  are formally represented by divergent series  in powers of the coupling, much like the expanded correlator itself. 

\begin{figure}\begin{center}
\includegraphics[width=6.5cm]{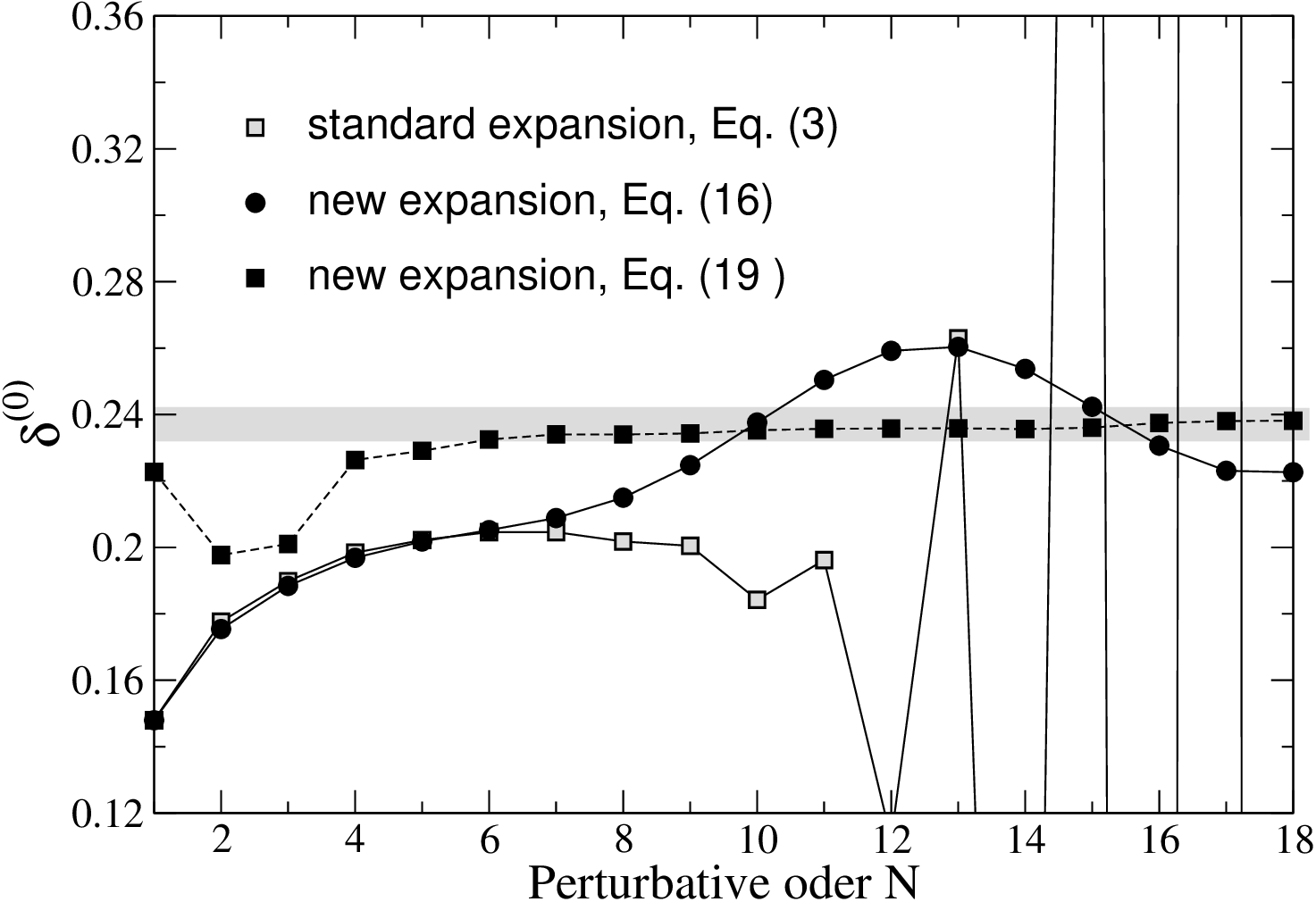}
\caption{\label{fig:CIstw} Values of $\delta^{(0)}$ for the model discussed in \cite{BeJa}, obtained with the standard and the new expansions,  as a function of the perturbative order $N$.   The horizontal band is the exact value from \cite{BeJa} with an experimental error. The calculations are performed with $a_s(m_\tau^2)=0.34/\pi$. }\end{center}
\end{figure}	

A detailed comparison of the standard and the new expansions  (\ref{DnewCI0}) and  (\ref{DnewCI}) was performed in \cite{CaFi3}, using  as reference a realistic model for the Adler function proposed in \cite{BeJa}. This model, which expresses the Borel transform in terms of a few singularities, allows the calculation of the exact Adler function and of its perturbative approximants. To illustrate the comparison, we shall use the contour integral
\be\label{delta0} \delta^{(0)} =  \frac{1}{2\pi i}\, \oint\limits_{|s|=s_0}\, \frac{d s}{s}\, \omega(s)\,\hat D(s)\,,
\ee
where $s_0=m_\tau^2$ and  $\omega(s)=1-2s/s_0 + 2(s/s_0)^3-(s/s_0)^4$ is  a kinematical factor. The  quantity  $\delta^{(0)}$  can be determined experimentally in terms of the total hadronic width of the $\tau$ lepton \cite{BNP92}. Therefore, the integral (\ref{delta0})  is of interest for the experimental determination of the strong coupling from the hadronic $\tau$-decays. 

To perform the comparison, we calculate  $\delta^{(0)}$ using the exact expression of the Adler function  in the model \cite{BeJa}, and its  expansions truncated at a finite order $N$. The running coupling $a_s(s)$  is obtained by integrating numerically the renormalization group equation (\ref{reneq}) step by step along the circle $|s|=m_\tau^2$, in terms of a reference value, taken  to be $a_s(m_\tau^2)$. This procedure, which avoids the large imaginary logarithms appearing in expansions  at a fixed scale  is known as ``contour improved'' expansion of    $\delta^{(0)}$  \cite{BNP92}.  For illustration we  use $a_s(m_\tau^2)=0.34/\pi$, as in \cite{BeJa, CaFi3}.

 The results are  shown in Fig. \ref{fig:CIstw}, where the gray band represents the exact result, to which we attached for 
convenience an uncertainty equal to the experimental error (see section \ref{sec:alphas}). As seen in   Fig. \ref{fig:CIstw},    the standard expansion is not able  to reproduce the exact value at low  truncation orders $N$. Moreover, for higher $N$ the results start to oscillate wildly, due to the divergent character of the series.
On the other hand, our new expansions  (\ref{DnewCI0}) and  (\ref{DnewCI}) reproduce more closely the exact value. The expansion  (\ref{DnewCI}), which includes explicitly the singular behaviour at the first singularities, gives very good values even for low $N$, while the expansion  (\ref{DnewCI0}) gives a good approximation only at larger truncation orders $N$. This is due to the fact that, in the model proposed in  \cite{BeJa},  the strength of the first singularities, expressed by the residues in (\ref{branch}), is quite important. Therefore,  a large number of terms are required in order to describe them, if the singularity is not explicitly factorized. In other models, with milder leading singularities, the expansion (\ref{DnewCI0}) approximates well the exact results even
 at low values of  $N$. Both expansions, (\ref{DnewCI}) and (\ref{DnewCI0}), are actually  much better than the standard expansion (\ref{DCI}) in powers of the coupling.

\vskip0.5cm
\section{OTHER ADMISSIBLE EXPANSIONS \label{sec:gen}}

As remarked above (see also \cite{CaFi, CaFi3}), while the expansion (\ref{Bw0}) is unique, the explicit inclusion of the first singularities of $B(u)$  contains some arbitrariness. The description of the singularities by multiplicative factors is a possibility, but is  not a priori necessary. Moreover, the factors are not unique.  For a large number of terms in the expansion the form of these factors is irrelevant, but at low orders one prescription may be better than another.  In (\ref{DnewCI})   the dominant behaviour was included by simple singular factors expressed in the $w$ variable. In this section we briefly discuss several other  possible expansions. A more complete analysis will be presented elsewhere.

To construct alternative approximants,  we first remark that  the dominant behaviour may be included by  singular factors expressed in the $u$ variable, instead of $w$. One may further multiply the leading factors by other functions analytic in the $u$-complex plane cut along the real axis for  $u \ge 2$ and $u \le- 1$. In particular, one may consider singularities on an unphysical Riemann sheet, or  placed at $u=3$ and $u=-2$, which are expected to occur as next-to-leading renormalons in the physical case.  The additional factors can be expressed either in the variable $u$ or in the variable $w$. We consider for illustration a set of new expansions written in the compact form 
\be\label{DnewCIk}
\hat D(s)=\sum\limits_{n\ge 0} f_{n,k}\, {\cal W}^k_n(s), 
\ee
where the expansion functions are defined as
\be\label{Wnk}
{\cal W}_n^k(s)=\frac{1}{\beta_0}{\rm PV} \int\limits_0^\infty\!{\rm e}^{-u/(\beta_0 a_s(s))} \,  \frac{ w^n}{{\cal F}_k(u)}\,{\rm d}u,
\ee
with the specific choices
\bea\label{F}
{\cal F}_1(u)\!\!&=&\!\!(1-u)^{\gamma_1}(1-u/2)^{\gamma_2},\\
{\cal F}_2(u)\!\!&=&\!\!(1+w)^{2\gamma_1}(1-w)^{2\gamma_2} (2-w)^2 (3-w)^2,\\
{\cal F}_3(u)\!\!&=&\!\!(1+w)^{2\gamma_1}(1-w)^{2\gamma_2}(1+u/2)^2(1-u/3)^2,\\
{\cal F}_4(u)\!\!&=&\!\!(1+w)^{2\gamma_1}(1-w)^{2\gamma_2} \prod_{j=1}^4\, \left(1-\frac{w}{w_k}\right)^2,  \\
{\cal F}_5(u)\!\!&=&\!\!(1+w)^{2\gamma_1}(1-w)^{2\gamma_2} \prod_{j=1}^4 \,\left(1-\frac{w}{w_k}\right)^{0.5}.  
\eea
Here $w\equiv w(u)$, where the function $w(u)$ is defined in (\ref{w}), and we denoted by $w_1=w(3+i\epsilon)$,  $w_2=w(3-i\epsilon)$,  $w_3=w(-2+i\epsilon)$,  $w_4=w(-2-i\epsilon)$  the positions the points $u=3 \pm i\epsilon $ and $u=-2 \pm i\epsilon$ on the circle $|w|=1$ in the $w$-plane.  In the absence of 
further information, the  exponents of the additional singularities  in  ${\cal F}_3(u)$, ${\cal F}_4(u)$ and  ${\cal F}_5(u)$ are arbitrary. 

\begin{figure}\begin{center}
\includegraphics[width=6.cm]{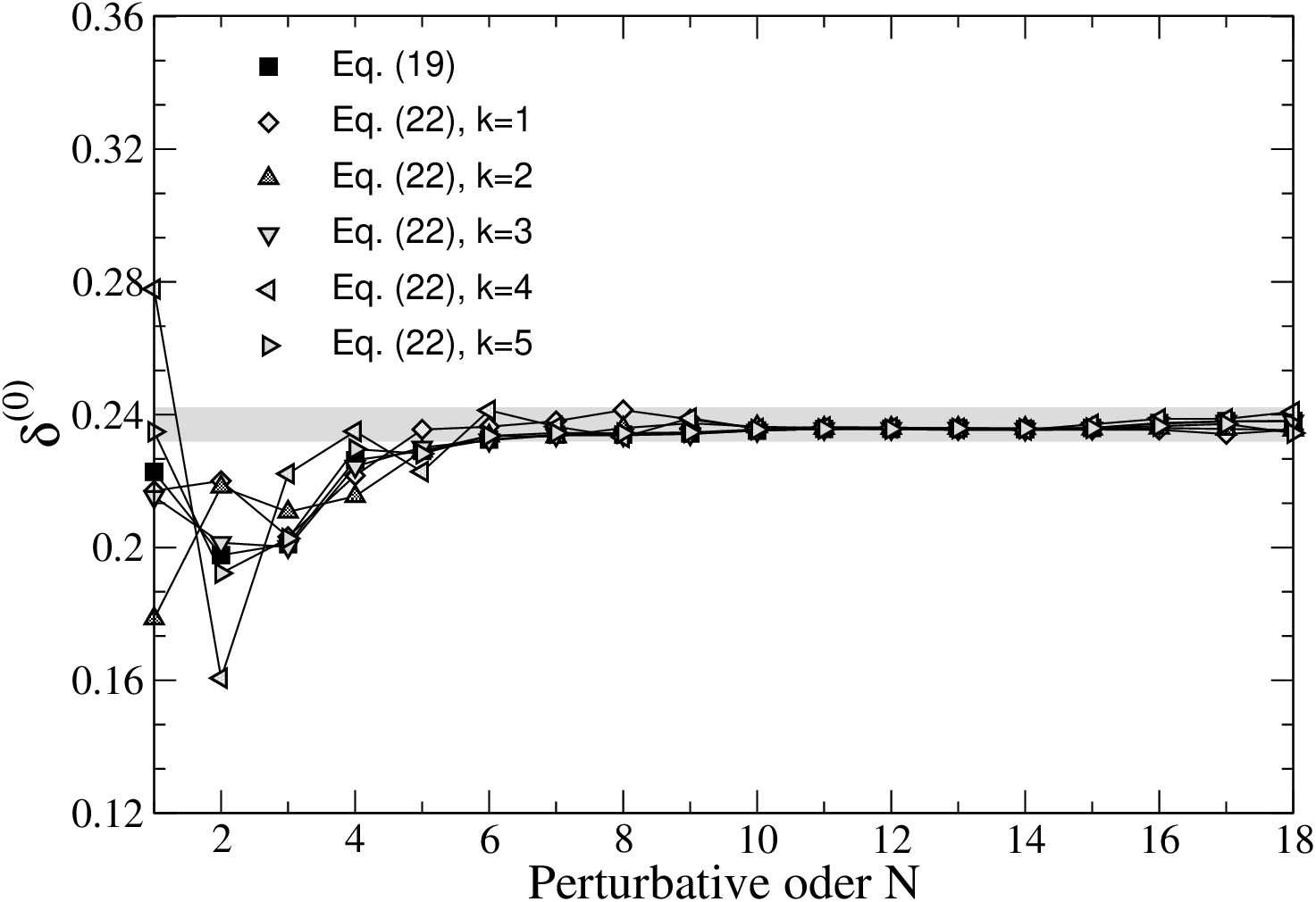}\hspace{1cm}\includegraphics[width=6.cm]{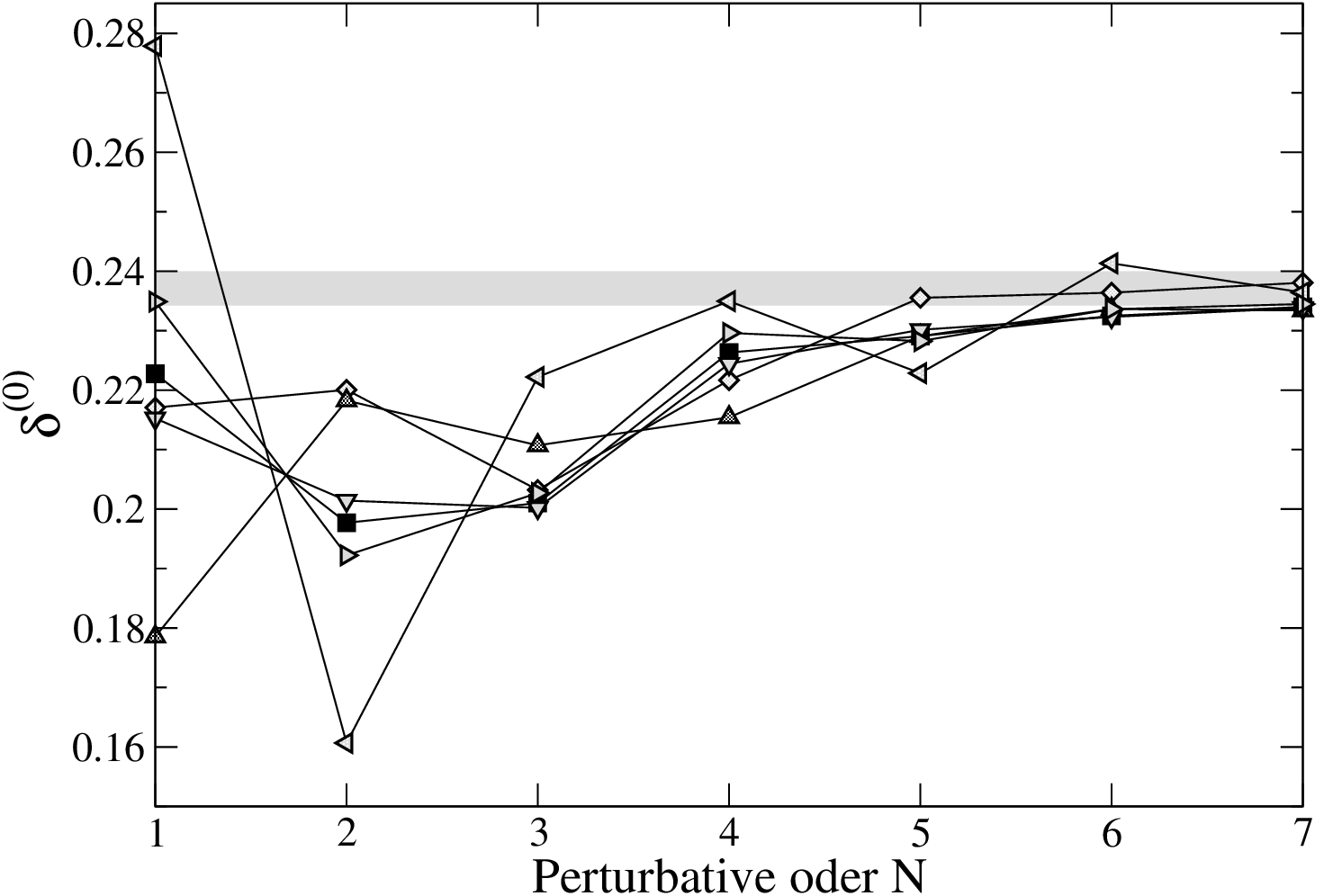}
\caption{\label{fig:CI} Values of $\delta^{(0)}$ for the model  \cite{BeJa}, obtained with various expansions, as a function of the perturbative order $N$.   The horizontal band is the exact value with an experimental error. The right panel enlarges
  the figure for  $N\le 7$. }\end{center}
\end{figure}

In Fig. \ref{fig:CI} we test the eficiency of the above approximants, using the  model \cite{BeJa}  and  $a_s(m_\tau^2)=0.34/\pi$ as in the previous section. The results shown in the left panel confirm that at large $N$ all the choices become equivalent, as expected\footnote{A possible divergency at still higher values of $N$, not excluded in principle, is  discussed in \cite{CaFi1, CaFi2, CaFi3}.}.  At low $N$ the differences between various approximations are more pronounced, but starting from $N=5$, which corresponds to the present status in QCD, the range becomes rather narrow. In particular, for $N=5$ the expansion  (\ref{DnewCI}) adopted in \cite{CaFi3} gives the central value, the spread around it being  $\pm 0.006$, close to the experimental uncertainty (we recall that for comparison we attached an "experimental" error to the exact result, as shown by the gray band in Fig. \ref{fig:CI}).   We shall  discuss in the next section the implications of these results on the determination of $\alpha_s$.

\vskip0.5cm
\section{DETERMINATION OF $\alpha_s$ \label{sec:alphas}}

The precise determination of $\alpha_s$ from the hadronic decays of the $\tau$ lepton is one of the most important results in perturbative QCD (for a recent review see \cite{Bethke}).  The problem was revisited  recently \cite{Davier2008, BeJa, CaFi3}, after the  calculation of the Adler function to four loops  \cite{BCK08}, the same order at which   the $\beta$ function of the renormalization group equation  is known \cite{LaRi, Czakon}. 

As mentioned in the previous section, the quantity  relevant for the extraction of $\alpha_s(m_\tau^2)$ is the integral (\ref{delta0}). We shall adopt the phenomenological value quoted in \cite{BeJa}:
\be\label{phen}
\delta^{(0)}_{\rm phen}=0.2042 \pm 0.0050.
\ee
The determination of $\alpha_s(m_\tau^2)$ then amounts to solving the equation
$\delta^{(0)}=\delta^{(0)}_{\rm phen}$, with $\delta^{(0)}$ calculated from  (\ref{delta0})  using various theoretical expansions of $\hat D(s)$. We use as input the known coefficients $K_n$ from (\ref{Ki}) and $K_5$ from (\ref{K5}), and the running coupling expressed in terms of $\alpha_s(m_\tau^2)$ by solving numerically the renormalization group equation (\ref{reneq}) along the integration contour.  

Taking the average over the various functional forms discussed in the previous sections, we obtain
\be\label{final}
\alpha_s(m_\tau^2)=  0.320 \pm 0.012\,.
\ee
where the  error includes  the experimental uncertainty quoted  in (\ref{phen}), the effect of a 50\% variation of the coefficient $K_5$, the uncertainty of renormalization scale  \cite{CaFi3}, and an additional error of $\pm 0.003$ due to the freedom in choosing  the expansion.

\vskip0.5cm
 
\section{CONCLUSIONS \label{sec:conc}}

The new perturbation expansion for QCD observables, proposed in \cite{CaFi}, is based on an optimal expansion of the Borel transform, which  reproduces  order by order the known perturbative coefficients calculated from Feynman diagrams and converges in the whole complex  $u$-plane cut along two lines of the real axis. While the optimal expansion variable $w$  is unique, a freedom  exists in the way  of including additional information, like the known behaviour  of the Borel transform near the first branch points. The inclusion of this behaviour is neither unique nor precisely formulated and we have no optimal method, in contradistinction to the choice of the optimal $w$, which is based on an ``optimality'' theorem  \cite{CiFi}.  

In the present paper we investigated a set of possible ways of implementing the behaviour of the Borel transform at the leading singularities  and tested their properties in the frame of a model proposed recently in \cite{BeJa}.  The results show that, for the number of perturbative terms calculated at present from Feynman diagrams in QCD, the various approximants give very consistent results.

Our analysis  shows that the freedom in the implementation of the known singular behaviour  of the Borel transform does not affect practically  the extraction of  $\alpha_s$ from $\tau$ hadronic decays: the central value of $\alpha_s(m_\tau^2)$ is unmodified, the only effect being
   an additional error on  of about $\pm 0.003$, comparable with the experimental error. Our final result is given in (\ref{final}). A  detailed analysis,  including a more general class of approximants and the investigation of alternative physical models for the QCD correlator, will be presented in a future work.  

\vskip0.5cm

\subsubsection*{Acknowledgements}
  This work was supported by CNCSIS in the Program Idei (Contract No. 464/2009),  and by the Projects No. LA08015 of the  Ministry of Education and AV0-Z10100502 of the Academy of Sciences of the Czech Republic.

\end{document}